\documentclass[twoside]{article}
\usepackage{fleqn,espcrc2,epsfig}

\usepackage{graphicx}

\newcommand{\beq}{\begin{equation}}
\newcommand{\eeq}{\end{equation}}
\newcommand{\bea}{\begin{eqnarray}}
\newcommand{\eea}{\end{eqnarray}}

\def\ta {\widetilde{\alpha}}

\def\mom{{_{\widetilde{\rm \kern -1pt M\kern -1pt O\kern -1pt M}}}}
\def\momt{$\widetilde{\rm\scriptstyle M\kern -1ptO\kern -1ptM}$ }
\def\ms{{_{\kern 1pt\overline{ \rm M \kern -1ptS}}} }
\def\mst{${\rm\scriptstyle \overline{ M\kern -1pt S}}$ }

\newcommand{\AmS}{{\protect\the\textfont2
  A\kern-.1667em\lower.5ex\hbox{M}\kern-.125emS}}

\title{Gluon propagator, triple gluon vertex and the QCD coupling constant 
\thanks{Poster presented  by Ph. Boucaud {\tt (phi@th.u-psud.fr)}\hfill\break 
$^\dagger$ Unit\'e Mixte de Recherche  UMR 8627 du CNRS\hfill\break
$^+$Unit\'e Mixte de Recherche C7644 du CNRS}
}
	
\author{ D. Becirevic
\address{Dip. di Fisica, INFN, Sezione di Roma, P.le Aldo Moro 2, I-00185 Rome, Italy.},
 Ph. Boucaud  
\address{Laboratoire de Physique Th\'eorique$^\dagger$,  B\^atiment 211,
Universit\'e de Paris XI,  91405 Orsay, France.},
 J.P. Leroy$\ \rm ^b$, 
J. Micheli$\ \rm ^b$,  O. P\`ene$\ \rm ^b$, J. Rodr\'\i guez--Quintero$\ \rm ^b$\\ 
and C. Roiesnel  
\address{Centre de Physique Th\'eorique de l'Ecole Polytechnique$^+$,
91128 Palaiseau Cedex, France.
 }
}

\begin{document}

\begin{abstract}
We study the UV-scaling of the flavorless gluon propagator
in the Landau gauge in an energy window up to 9 GeV. 
Dominant hypercubic lattice  artifacts are eliminated.
A large set of renormalization schemes is used to test asymptotic scaling. 
We compare with our results obtained directly from the triple gluon vertex.
We end-up with $\Lambda_{\overline{\rm{MS}}} \ = \ 318(12)(5)$ MeV
and 292(5)(15) MeV respectively for these two methods, compatible which each other but
significantly above the Schr\"odinger method estimate.
\vspace{1pc}
\end{abstract}

\maketitle

\section{Definition of the coupling constant from the propagator}
The Euclidean two point Green function in momentum space writes in the
 Landau gauge: 
\beq
	G_{\mu\nu}^{(2)\,a b}(p,-p)=G^{(2)}(p^2) 
	\delta_{a b} \left(\delta_{\mu\nu}-
	\frac{p_{\mu}p_{\nu}}{p^2}\right)\label{G2}
\eeq
where $a, b$ are the color indices.

In any regularization scheme (lattice, dimensional regularization, etc.)
with a cut-off $\Lambda$ ($a^{-1}$, $(d-4)^{-1}$, ...)
the bare gluon propagator in the Landau gauge is such that
\beq
\Gamma(p^2) \ \equiv \ \lim_{\Lambda\to \infty}
\frac{d \ln [p^2 G^{(2)}_{\rm bare}(p^2,\Lambda)]}{d \ln p^2},
\label{obs}\eeq
is independent of the regularization scheme.

Lattice calculations provide us with a 
measurement of the bare propagator, and hence of the observable $\Gamma$
in eq. (\ref{obs}).

We { \bf define} the running coupling constant, $\ta(p^2)$, through:
\beq
\Gamma(p^2)  \ =  \ 
-\frac{\gamma_0}{4\pi} \ \ta
-\frac{\widetilde\gamma_1} {(4\pi)^2} \ \ta^2
-\frac{\widetilde\gamma_2} {(4\pi)^3} \ \ta^3 \label{gamma}
\eeq
for a given set of real coefficients $\gamma_0$, 
$\widetilde\gamma_1$ and $\widetilde\gamma_2$.

Note that in the MOM scheme, the gluon renormalization
constant $Z_3$ is given by 
 $ Z_3(q^2, \Lambda) = q^2 G_{bare}(q^2,\Lambda)$.
Our observable $\Gamma$ is then the anomalous dimension of the gluon
renormalization constant in the MOM scheme.
But it is important to realize that this does not imply that 
we are  stuck to the MOM scheme :
 $\Gamma$ is defined in a scheme independent way by eq.(\ref{obs})
 and can be  studied in any general scheme defined by eq.(\ref{gamma}).
Nevertheless this imposes the value of $\gamma_0$ to be equal
to  13/2 , the first (universal) term of the anomalous dimension.

Our strategy to study  $\widetilde\alpha(\mu)$  on the lattice
from the propagator is the following:\hfill\break
$\bullet$ Measure on the lattice 
     the bare gluon propagator in Landau gauge, extract
      $ G^{(2)}_{\rm bare}(p,a)$ 
       following eq.(\ref{G2}) and compute
      $\ln\left[ Z_3(p,a)\right] 
     = \ln \left[p^2 G^{(2)}_{\rm bare}(p,a)\right]$.
\hfill\break
$\bullet$  Choose $\widetilde\gamma_1$, $\widetilde\gamma_2$ 
      and integrate the two coupled differential equations:
    \beq
     \frac{d \ln Z_3(q,a)}{d \ln q^2}
       =  - \frac{\gamma_0}
     {4\pi}\ \ta -\frac{\widetilde\gamma_1} {(4\pi)^2}\  \ta^2
     -\frac{\widetilde\gamma_2} {(4\pi)^3} \ \ta^3
     \eeq
     \beq
     \frac{\partial \widetilde\alpha}{\partial \ln q}
     = -\frac{\beta_0}
     {2\pi} \widetilde\alpha^2 -\frac{\beta_1} {(2\pi)^2} \widetilde\alpha^3
     -\frac{\widetilde\beta_2} {(4\pi)^3} \widetilde\alpha^4 
     + O(\widetilde\alpha^5)\nonumber\label{beta}
     \eeq     
     ($\widetilde\beta_2$ is known for any $\widetilde\gamma_1$ 
     and  $\widetilde\gamma_2$\cite{beta1}).
     The solution for $Z_3(q,a)$ and $\widetilde\alpha(q)$ 
      depends  on the values $Z_3(\mu,a)$ 
     and $\widetilde\alpha(\mu)$ given at some initial value 
      $q=\mu$.
\hfill\break
$\bullet$  Fit the lattice data with the previous solutions to determine
        the "best"  $Z_3(\mu,a)$ and $\widetilde\alpha(\mu)$.
	Extract $\widetilde\Lambda$ (we use  three-loop expressions here)
	and convert to $\Lambda_{\overline{\rm MS}}$ through
$ \Lambda_{\overline{\rm MS}} = \widetilde \Lambda \,\exp\left[{\frac
 {\widetilde\gamma_1-\overline\gamma_1 } 
 {2 \gamma_0\beta_0}}\right]$ where $\overline\gamma_1 \simeq 155.3$ is 
 the value of the coefficient
 $\widetilde\gamma_1$ in the $\overline{\rm MS}$ scheme. 
\section{Hypercubic artifacts}
In an hypercubic volume the momenta are the discrete
sets of vectors $p_\mu = \frac{2 \pi} {a L} n_\mu$ where the $n_\mu$ are
integer. For a given orbit of the continuum isometry group
(characterized by a given value for  $n^2$), there are  distinct orbits
of the hypercubic isometry group. For example $n_\mu$ = (1,1,1,1) and 
 $n_\mu$ = (2,0,0,0).

A usual approach to reduce these hypercubic effects is to use
only the "democratic" sets of momenta, i.e. those for which the
momentum is equally distributed on the different directions  
and/or the use of the lattice momenta 
$\widetilde p_\mu \equiv \frac {2} {a} \sin\left( \frac { a p_\mu}{2}\right)$.
We propose another approach, thanks to our large statistics, which allows
us to eliminate  these hypercubic artifacts.
On the lattice, an invariant scalar form factor  like $G^{(2)}$
is indeed a function of the 4 invariants~: $p^{[m]} \equiv \sum_\mu p_\mu^m$
with $m = 2,4,6,8$.
We will neglect the invariants with degree higher than 4 since they vanish at
least  as $a^4$.

When several orbits exist for one $p^2$, 
we have found a nice linear behavior
 $G_{\rm lat}^{(2)}(p^2,a^2 p^{[4]}) 
= G_{\rm lat}^{(2)}(p^2,0) \ + \ \left. 
{\partial G\over a^2 \partial p^{[4]}} 
\right|_{a^2 p^{[4]}=0} a^2 p^{[4]} $.
It is then possible to extrapolate
to $p^{[4]}=0$ and we can use  $G^{(2)}_{bare}(p^2) =  \lim_{p^{[4]}\to 0}
G_{\rm lat}^{(2)}(p^2,p^{[4]}) $  in all our analysis.
We have introduced this method in \cite{beta3}. 
To reach higher energies and exploit also the cases where only one 
orbit is available, the method has been  extended~:
from dimensional argument, the slope 
$\frac {\partial G } { a^2 \partial p^{[4]} } $ is expected to 
behave like $ 1/p^4$ as $L \rightarrow \infty$. 
As shown in  \cite{beta4}, we found that the ansatz 
$
\left. {\partial G\over a^2 \partial p^{[4]}} 
\right|_{a^2 p^{[4]}=0} = \ {b\over p^4} \left(1+
c \ 
\exp(-d \ L  p) \right) 
$
gives a good description at fixed $\beta$ 
of all our data thus allowing us to extrapolate 
to $p^{[4]}=0$.

\section{Finite volume effects}

\begin{figure}[hbt]
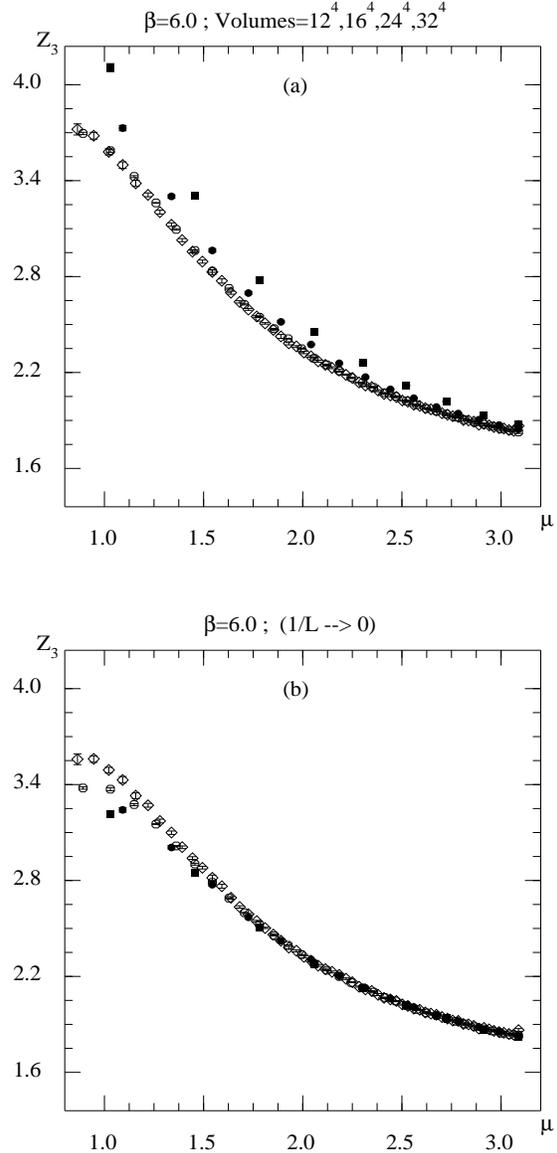

\begin{center}
\leavevmode
\epsfysize=8.0truecm
\epsfxsize=8.0truecm
\epsffile{vol60.eps}
\epsfxsize=8.0truecm
\epsfysize=8.0truecm
\epsffile{vol60I.eps}
\end{center}
\vspace*{-1.5cm}
\caption{\it Plot (a) shows the  free-hypercubic propagators 
evaluated on $12^4$ (black
squares), $16^4$ (black circles), $24^4$ (white circles), 
$32^4$ (white squares) lattices at 
$\beta=6.0$. 
Plot (b) shows the extrapolation $L \to \infty$ of these 
 free-hypercubic 
propagators  using the parametrization given in 
Eq. (\ref{vol}).}
\label{figvol}
\end{figure}

In \cite{beta3}, with our  lattices at $\beta=6.2$ and 6.4, 
we found that {\bf the gluon propagator has not yet reach asymptotic
scaling  at three-loop even at $\mu \simeq 5$ GeV}.
Higher energies are then required to test scaling. 
We have now lattices at $\beta=6.8$, but as we are limited to $24^4$
which is not a too large volume at this value of $\beta$,
 attention has to be paid to the finite volume effects.\hfill\break
 To that end we have generated  lattices at $\beta=6.0$ with several
   volumes $12^4$,$16^4$, $24^4$ and $32^4$
   and tried several functional forms to parametrize   the finite volume
   effects at fixed value of $\beta$.  Finally we proposed:
\bea
G_{\rm lat}^{(2)}(p^2,L,a) &=& G_{\rm lat}^{(2)}(p^2,\infty,a)\label{vol} \\ 
&\times &\left(  1+v_1 \left({a\over
L} \right)^4+v_2 \ {\rm e}^{-v_3 Lp}\right) \nonumber
\eea
which gives a good parametrization of all the $\beta=6.0$ data
in the UV part.  See figure (\ref{figvol}).

Once  the parameters
$v_1,v_2,v_3$  have been determined from the lattices
at $\beta=6.0$, the parametrization (\ref{vol}) can 
be applied to extrapolate our results on $16^4$ and $24^4$ lattices 
at $\beta=6.8$. 
The good agreement shown by the  curves resulting from the extrapolation 
(figures can be found in \cite{beta4}) is a crucial test
for the validity of such a parametrization 
for the finite volume effects.

\section{Results}

 To calibrate the lattice at $\beta=6.2$, we have used the value for the the lattice
spacing which has  been measured recently  with a non-perturbatively 
improved action (free from {O(a)} artifacts) \cite{ape}. The measured  value is~: 
$a^{-1}(\beta=6.2)\ =\ 2.72
(11)$ GeV.  Other lattices have been calibrated
relatively to the one at $\beta = 6.2$ with the results for  
$a \sqrt{\sigma}$, the string tension in lattice units, published in \cite{bali}.
We took: $a^{-1}(\beta=6.0) = 1.94$ GeV, 
$a^{-1}(\beta=6.2) = 2.72$ GeV , $a^{-1}(\beta=6.4) = 3.62$ GeV
and $a^{-1}(\beta=6.8) = 6.03$ GeV.

We have applied the strategy described above to study 
$\widetilde\alpha$,
using the data at  $\beta=6.8$  and found 
at three-loop using the 
 ${\widetilde {\rm MOM}}\ $  scheme:
$\Lambda^{(3loop)}_{\overline {\rm MS}}\simeq 0.346\; 
\Lambda^{(3loop)}_{\widetilde {\rm MOM}} = 315 \pm 12 \ {\rm MeV}
$,
where the error is only statistical here.

We can exploit the large scheme space to which we have easily access
 to define a domain of ``good schemes" by imposing some
constraints on $\widetilde\gamma_1$ and $\widetilde\gamma_2$. 
The idea is
to  select schemes for which the successive terms in the  expansions in 
 $\widetilde\alpha$ for the 
$\beta$-function and $\Gamma$ are not exceedingly large compared 
to the previous terms 
(see \cite{beta3} for the precise definition of the cuts we used).
A large scheme dependence in this set of schemes 
would indicate that we are still far from 
perturbative scaling.
We found  that   \momt  
is a ``good scheme" while \mst {\bf is not} and the result for 
$\ \ \Lambda^{(3loop)}_{\overline {\rm MS}}$ ranges from 313 to 323 MeV
in the domain of ``good schemes". We consider this  dispersion to give an estimate
of the effect of higher order terms in the perturbative expansions and
 the flavorless   $\Lambda_{\overline {\rm MS}}$ 
  extracted from the propagator   up to 9 GeV we  quote is then:
\bea
\Lambda^{(propag.)}_{\overline {\rm MS}} = 318 \ (12) \ (5) \ 
\frac{a^{-1}(\beta=6.8)}{6.03\  {\rm GeV}}
{\rm MeV}
\label{lambda1}
\eea

We can compare  this result 
 to the one we obtain by 
direct standard textbook
methods from the triple gluon vertex in the \momt scheme \cite{beta1}.
Data for $\alpha$ are less
precise than those for the propagator so we have not yet tried to eliminate
the finite volume effects with a method similar to the one used here.
Consequently we did not use our $\beta=6.8$ lattice and stay with our large
volume lattices at smaller $\beta$.
$\Lambda$ was found to be nearly independent of $\beta$ for
$\beta=6.0,6.2$ and 6.4  \cite{beta1} :
\bea
\Lambda^{(vertex)}_{\overline {\rm MS}} = 292 \ (5) \ (15) \ {\rm MeV}
\frac{a^{-1}(\beta=6.8)}{6.03\  {\rm GeV}}
\label{lambda2}\eea
Our two methods, based on Green functions, give compatible results, 
  both significantly larger than the one obtained 
 over  a large energy range 
with the Schr\"odinger functional \cite{sommer}.

\end{document}